\documentclass[preprint,pre,floats,aps,amsmath,amssymb,superscriptaddress, nofootinbib,tightenlines, showkeys,showpacs]{revtex4-1}

\usepackage{amsmath,amssymb,amsfonts,amsthm}
\usepackage[english]{babel}
\usepackage{xcolor}
\usepackage[utf8]{inputenc}
\usepackage{graphicx}
\usepackage{eurosym}

\newtheorem{theorem}{Theorem}
\newtheorem{remark}[theorem]{Remark}
\newtheorem{proposition}[theorem]{Proposition}

\begin{document}

\markboth{F. Corvino, F. Flandoli, M. Leocata, G. Livieri, S. Morlacchi, A. Pirni}{Multiscale modeling of Green Energy Transition: Structural properties and an example}

\title{Structural properties in the diffusion of the solar photovoltaic in Italy: individual people/householder vs firms}

\author{Franco~Flandoli} 
\email{franco.flandoli@sns.it}
\affiliation{Scuola Normale Superiore, Pisa, Italy.}

\author{Marta~Leocata}
\email{mleocata@luiss.it}
\affiliation{, LUISS Guido Carli University, Rome, Italy.}

\author{Giulia~Livieri}
\email{g.livieri@lse.ac.uk}
\affiliation{The London School of Economics and Political Science, London, United Kingdom.}

\author{Silvia~Morlacchi}
\email{silvia.morlacchi@dm.unipi.it}
\affiliation{Dipartimento di Matematica, Università di Pisa, Pisa, Italy.}

\author{Fausto~Corvino}
\email{fausto.corvino@gu.se}
\affiliation{University of Gothenburg, Gothenburg, Sweden.}

\author{Alberto~Pirni}
\email{alberto.pirni@santannapisa.it}
\affiliation{Scuola Superiore Sant’Anna, Pisa, Italy.}

\date{\today}
\begin{abstract}
This paper develops two mathematical models to understand subjects' behavior in response to the urgency of a change and inputs from governments \textcolor{black}{e.g., (subsides)} in the context of \textcolor{black}{the diffusion of the solar photovoltaic in Italy}. The first model is a Markov model of interacting particle systems. The second one, \textcolor{black}{instead}, is a Mean-Field Game model. In both cases, we derive the scaling limit deterministic dynamics, \textcolor{black}{and we} compare the \textcolor{black}{latter} to the Italian solar photovoltaic data. We identify periods where the first model describes the behavior of domestic data well and a period where the second model captures a particular feature of data corresponding to companies. The comprehensive analysis, integrated with a philosophical inquiry focusing on the conceptual vocabulary and correlative implications, leads to the formulation of hypotheses about the efficacy of different forms of governmental subsidies. 
\end{abstract}

{
\let\clearpage\relax
\maketitle
\begin{small}
\end{small}
\vfill
}
\section{Introduction}\label{sec:introduction}
Green Energy Transition (GET, henceforth) has, \textcolor{black}{in general}, many facets. One is the problem of understanding subjects' reactions to the urgency of a change and inputs from governments (e.g., subsidies). The latter is a fascinating question at the intersection of several cultural fields. This paper investigates this interrogative through the lens of mathematicians and philosophers by examining a specific \textcolor{black}{case study}, \textcolor{black}{namely the diffusion of} solar photovoltaic panels \textcolor{black}{in Italy}, to identify concrete issues and have some data comparing our proposed models; \textcolor{black}{Section \ref{sect development Italy} in Appendix gives a short description of the development of solar photovoltaic in Italy.} \textcolor{black}{More precisely, in the present paper, the term ``subjects" indicates either individual people and householders or companies. The purpose is to understand which type of structural modeling -- to be detailed below -- better captures the behavior of these subjects in order to formulate a hypothesis about the efficacy of different forms of governmental subsidies. We stress right away that we are not interested in investigating the ability of models to simulate emerging behaviors, and we postpone this latter problem to future research. Here, the validation of models consists of the latter's capability to fit specific empirical data. In particular, we construct two mathematical models aiming to describe different (structural) features of the considered GET and see the consequences of these choices; we now describe the characteristics of these models from a behavioral and a mathematical perspective.}\\ \\
\indent     From a behavioral perspective, the first model comprises \textcolor{black}{subjects} with a clear orientation who favor installing solar photovoltaic panels but often do not complete the action; individual people or householders inspire this model. The subjects' assessment, influenced by, e.g.,  personal experiences, perceived social norms, and beliefs about the technology's benefits and drawbacks, of the likelihood of various outcomes associated with adopting solar PV, is related to what is known in the literature as \emph{subjective probabilities}; see, e.g., \cite{suppes1969role}. \textcolor{black}{In particular, several studies have found that many householders do not invest in energy-efficient technologies, even if cost-benefit analyses have consistently shown that such technologies provide a high internal rate of return and are socially beneficial (e.g., \cite{chapman2009retrofitting, clinch2001cost, webber2015impacts}). Reasons for the just-mentioned energy-efficiency gap comprise investment inefficiencies and behavioral factors, such as imperfect information, uncertainty about the benefits of the investments, and decision-making biases. One of these decision-making biases is the so-called ``bounded rationality". Human beings are bounded rational: they have limited cognitive resources, which naturally constrains optimal decision-making; see, e.g., \cite{frederiks2015household}. In particular, empirical evidence from psychology and behavioral economics shows that consumer choices and actions often deviate systematically from the neoclassical economic assumptions of rationality, and certain fundamental and persistent biases in human decision-making regularly produce behavior that these assumptions cannot account for; see, e.g., \cite{wilson2007models, pollitt2013role}. There exist a plethora of cognitive biases and behavioral anomalies influencing consumer's patterns of energy usage, with the \textit{status quo bias}, \textit{loss and risk aversion}, \textit{sunk-cost effects}, \textit{temporal and spatial discounting}, and the \textit{availability bias} being the most powerful and pervasive ones; see  \cite{frederiks2015household}, Page 1386, for a nice description. Because a comprehensive review and modeling of these cognitive biases and behavioral anomalies are beyond the scope of the present paper, for the sake of presentation only, we will speak about \textit{procrastination}, a behavioural bias that occurs in the subsequent situation. Perceive things as less valuable or significant if further away in time (\textit{temporal discounting}) or space (\textit{spatial discounting}), even if such things afford long-term benefits. For instance, many people prefer $\$$100 in six years rather than $\$$50 in four years -- the discount rate of future utility, in other words, makes the value of $\$$100 two years later lower than the value of $\$$50 two years earlier, but not so lower as to be lower than the value of
$\$$50 two years earlier. As many people, however, prefer $\$$50 today to $\$$100 two years from now. This means that the choice between the two options is not only influenced by the utility units and their temporal detachment, but also by how close the option with the closer gains is to us. This tendency to be short-sighted and make time-inconsistent judgements often leads indeed to \textit{procrastination}\footnote{\textcolor{black}{Notice that procrastination is distinguished from two other psychological phenomena: hypocrisy and akrasia.}}, inertia and decreased cooperation in group settings; see, e.g., \cite{jacquet2013intra}. In particular, we summarize the previous cognitive biases in a state named \textit{Deliberating}, which indicates individual people or householders that can articulate a complex process lying behind the assumption of a specific pattern of action, namely the solar panels' installation, but that they have not done the latter action yet. The second model, instead, comprises subjects who make more \textit{rational decisions} based on optimization rules. This model is more inspired by the behavior of companies/firms. Indeed, some existing research works in the literature treat firms as economic agents with a rational maximization behavior; among these, we cite here \cite{judd1986dynamic}, \cite{berck1988dynamic}, \cite{gotoh2021framework}, \cite{semmler2022limit}, and \cite{chan2017fracking}.}\\
\indent Despite these differences, both models hinge on the \textcolor{black}{existence of an \textit{interaction} among subjects.} The present paper summarizes this interaction as dependence on specific rules of the mathematical models on the \textit{fraction of subjects that have already installed the solar photovoltaic panels at time $t$}, out of a population of size $N$; \textcolor{black}{in order not to introduce too many concepts here, only in this section, we will denote the previous quantity by $\frac{N_{G}\left(t\right)}{N}$.} How the term $\frac{N_{G}\left(t\right)}{N}$ enters is very different between the two models. More precisely, \textcolor{black}{in the first model}, the motivation for introducing $\frac{N_{G}\left(t\right)}{N}$ is a sort of \textit{imitation}, the orientation to participate more confidently in the GET when we observe that more people have done it. \textcolor{black}{Indeed, attitudes and behaviours of others influences people, which tend to follow norms reflecting what is socially approved (i.e., injunctive norms, which motivate by providing social rewards/punishment) and/or common (i.e., descriptive norms, which motivate by providing suggestions about effective and adaptive behaviour); see, e.g., \cite{feldman1984development, cialdini1998social}. In the second model, instead, there are subsidies to participate in the GET with a limited global amount, and when more \textcolor{black}{companies} have installed the panel, less accessible to get the subsidy.}\\ \\
\indent From a mathematical point of view, in the first model, we end up with an interacting particle system of Markov type, in which the randomness corresponds to the uncertainty that an individual installs solar photovoltaic panels due to \textcolor{black}{bounded rationality}. We analyze its scaling limit when the number of subjects goes to infinity and get certain deterministic differential equations. \textcolor{black}{Notice that this type of modeling is also confirmed by the extensive literature on opinion dynamics in which the evolution of opinions in society is modeled through Markov chains; see, e.g., \cite{holley1975ergodic, lewenstein1992statistical, sirbu2017opinion}.} In the second model, we end up with a \textit{game}, precisely a Mean-Field Game (MFG, henceforth) since it is an optimization problem where certain elements depend on the global mean quantity $\frac{N_{G}\left(t\right)  }{N}$. \textcolor{black}{We defer the interested reader to the nice books of \cite{carmona2018probabilistic} for a presentation of the MFG theory.} Here, we get a backward-forward system of differential equations in the scaling limit. \textcolor{black}{MFG models linked to GET can be found in, e.g., \cite{aid2021entry, dumitrescu2024energy}, and references therein.}\\
\indent We compare the two models -- precisely, their scaling limits deterministic equations -- with a fit of the parameters on data of solar photovoltaic taken from~\cite{GSE}. It is essential to observe that the solar photovoltaic cumulative capacity of the Italian time-series is the sum of four time-series, corresponding to ``Domestic", ``Industry", ``Services", and ``Agriculture" sectors. We focus on the ``Domestic" and the ``Industry" time-series, where our individual-based modeling looks more appropriate (together, they represent around 70$\%$ of the total, so their interpretation is a substantial problem). These time-series immediately display a subdivision into two periods, separated roughly speaking by the year 2012. \textcolor{black}{Before this year, Italy had a firm subsidy policy, the so-called Feed-in-Tariff (FiT, henceforth), which was of moderate-size before 2011 and of substantial size around 2011. After 2012, the firm subsidy policy was weaker than the one before 2012.} In particular, both the time-series of ``Domestic" and ``Industry" show, roughly speaking, a relatively exponential solid increase (clearly identified by linear trends on a logarithmic scale) in the period before 2012 and a weaker exponential increase after 2012. In addition, the ``Industry" time-series had a linear increase around 2011, essentially absent in the ``Domestic" time-series. \textcolor{black}{Therefore}, we find it interesting to explain, by mathematical models, the following two issues:
\begin{itemize}
\item[(I)]  the periods of exponential increase;
\item[(II)] the single period of a linear trend, with a large derivative, around 2011,
for companies.
\end{itemize}
The Markov model easily fits the exponential increase periods. Still, it is not natural for an explanation of the solid linear increase of 2011. On the contrary, the MFG model, totally unsuitable for the exponential increase periods, explains well the linear increase of 2011 observed for companies. This empirical evidence leads to the following conclusion. Companies, around 2011, underwent a \textit{game}, in contrast to all other periods and \textcolor{black}{individual} people or householders. The planning ability of companies is superior to that of domestic ones. Hence, using control theory arguments to explain company behaviour is more natural. Nevertheless, the cost function depended on $\frac{N_{G}\left(  t\right)  }{N}$ because the global amount of subsidies was limited. Therefore, the structure of the problem is a game, not simply of control.\\
\indent It is important to emphasize a structural element of the game (or control) theory, opposite to a Markov model: the presence of a ``time zero'', an initial instant of time. In the Markov model, any time is a new starting time due to memory loss. In a game or control problem, there is a functional to be minimized: at time zero, the agent should plan the future activity to minimize the functional. This planning is not repeated as in a Markov model that continuously restarts. So, why 2011? \textcolor{black}{As said above and by} looking at the documents~\cite{GSE}, 2011 came after some years of moderate-size FiT subsidies, which prepared the ground and alerted the companies. In 2011, Italy proposed a much more substantial FiT subsidy. Companies were prepared and acted in a game, with the initial time of the subsidy call as the initial time of the game. Householders were not prepared; slowly, they reacted but not with the game's logic. \textcolor{black}{Also, it is essential to make clear that we do not propose two
models for the same period and the same type of subjects, but we propose the idea that around the year 2011 the “Industry” compartment, having a more rational attitude towards investment with respect the other compartments, entered the modality of a game, while the others did not. Admittedly, the boundary is very vague since we always live in a game and outside a game, depending on how much we stress certain factors. In particular, what we are claiming is that around the
year 2011 and for the “Industry” compartment the behaviour is better described by
a game. In contrast, outside that period, for the other compartments and the “Industry” compartment
the game logic is not the driving force as far as the data shows. Nevertheless, some fraction of game dynamics always exists. Finally, we would like to stress that a subject can decide to enter a game in a certain period of her/his/its life but stay outside in others; being part of a game is a decision anyone can take. In particular, we notice that there is nothing ad-hoc for using different models for different periods, although this approach is not orthodox since it is not coherent with the modelling approach usually adopted in the literature. To insist on this element, one possibility would be to introduce an extra jump process in our model with values in $\{0, 1\}$ with a rate. A value 0 means that we are outside the game logic but that with a probability rate, the process may jump to the game logic state 1 and remain there until another rate moves it back to zero. In other words, the subject behaves over her/his/its entire life in some way, and this way contains the probability of making the transition of being more or less a player and the rate of transition can be governed by an external input, like the number of subsidies. Put in this way, it is a single model, where being within the game or not is just an internal part of the state. Another possibility would be to consider the so-called regime switching modelling approach. However, the period of time we are considering is so short and the occurrence of the game state is so rare that this over-structure looks exaggerate.}\\ \\
\indent The solar photovoltaic, as an example of GET has
been investigated by many authors in the literature through the lenses of different approaches. We here mention the following works, which do not represent a comprehensive list. (i) The agent-bases approach of, e.g.,~\cite{Zhao}, \cite{Palmer} and \cite{Peralta}. The Agent-based approach offers a framework to explicitly model the adoption decision process of the agent of a heterogeneous social system based on their individual preferences, behavioral rules, and interaction/communication within a social network. (ii) Versions of
the popular Bass model (\cite{Bass}); see, e.g.,~\cite{DaSilva}, which state that the knowledge about such a system highly influences the diffusion of solar photovoltaic systems in Brazil. (iii) Finite element methods to
account for spatial heterogeneity; see, e.g.,~\cite{Kara}, which applies the finite element methods to forecast the diffusion of solar photovoltaic systems in southern Germany. (iv) Survey-based analysis; see, e.g.,~\cite{Colasante}. Interestingly, some references in the literature analyze whether localized imitation drives the adoption of photovoltaic systems (a phenomenon captured by the ratio $\frac{N_{G}\left(  t\right)  }{N}$ in our model); see, e.g.,~\cite{Rode}, \cite{Baranzini}, \cite{Chad}, \cite{Copiello}, \cite{Curtius},
among (many) others. This issue and modeling related to interacting agents can be found in more general literature, like~\cite{marketing}. However, we have not found a comparison between a Markov modeling
and a MFG one to understand certain
differences observed in solar photovoltaic data.\\ \\
\textcolor{black}{\indent The paper is organized as follows. In Section \ref{sect time-series}, we describe some data which motivate the mathematical models presented in Section \ref{sec:Markov} (the Markov model) and \ref{sec:Meanfieldgameapproach} (the MFG model). Section \ref{sect Conclusions} concludes.}

\section{Examples of time-series\label{sect time-series}}
\begin{figure}[pb]
\centerline{\includegraphics[width=3.8in]{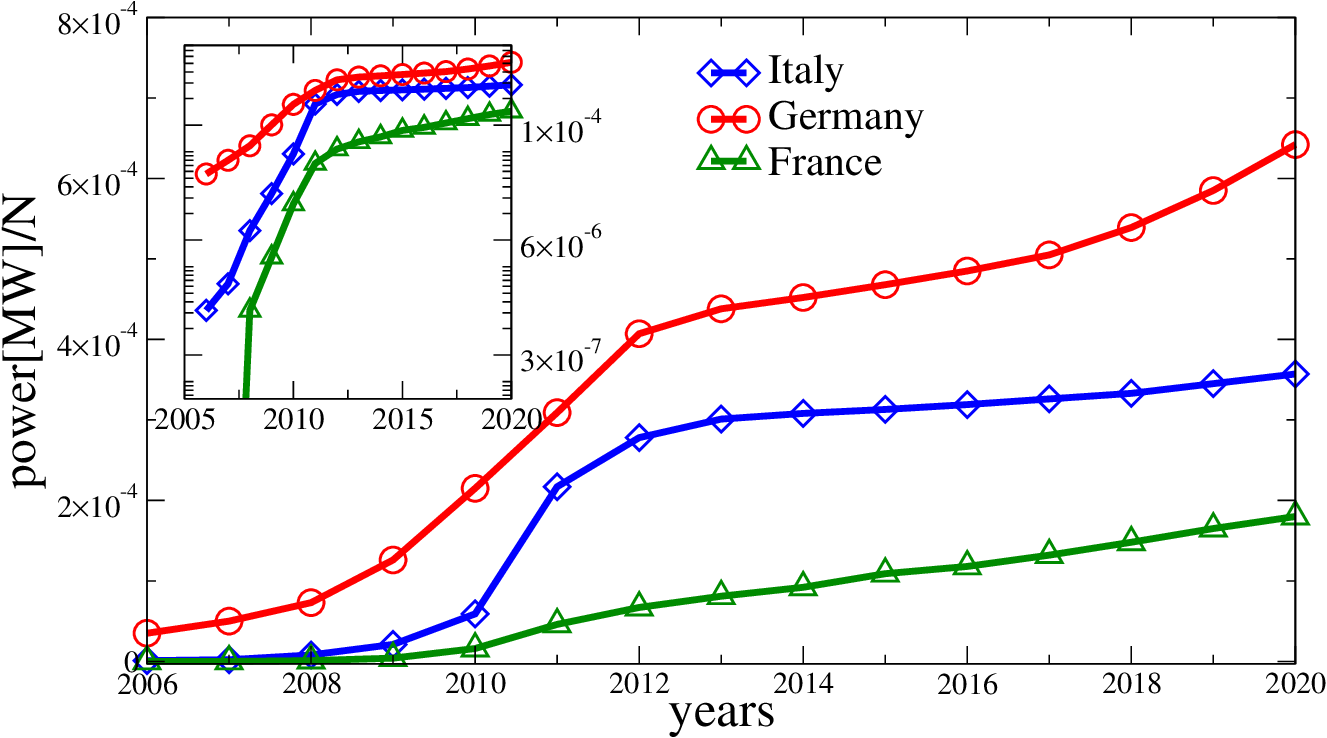}}
\vspace*{8pt}
\caption{Global photovoltaic power in Italy, Germany and France.}\label{fig:nations}
\end{figure}
\textcolor{black}{To gain a better understanding visually, let us start by looking at the time series of Italy, which is our main focus, and also Germany and France for a brief comparison. This will cover the period from 2006 to 2020 and show the overall power of photovoltaic installation without any specific categorization. Figure \ref{fig:nations} shows the three time series in both linear and logarithmic scales. One common observation is the change of regime around 2012, as discussed in Section \ref{sect development Italy}. Again, we notice that the term \textit{change of regime} must be understood as the presence of strong and innovative incentives opening the possibility of a game logic, possibility that would not have occurred without such incentives. The second common aspect is the approximate exponential growth in each of the two periods, which is more evident in the logarithmic scale. The focus on Figure \ref{fig:nations} is to emphasize the similarities, indicating that a mathematical model based on a two-period exponential growth is natural. The differences between the three countries are not the focus here, except for some comments in the conclusions section.}

\begin{figure}[pb]
\centerline{\includegraphics[width=3.8in]{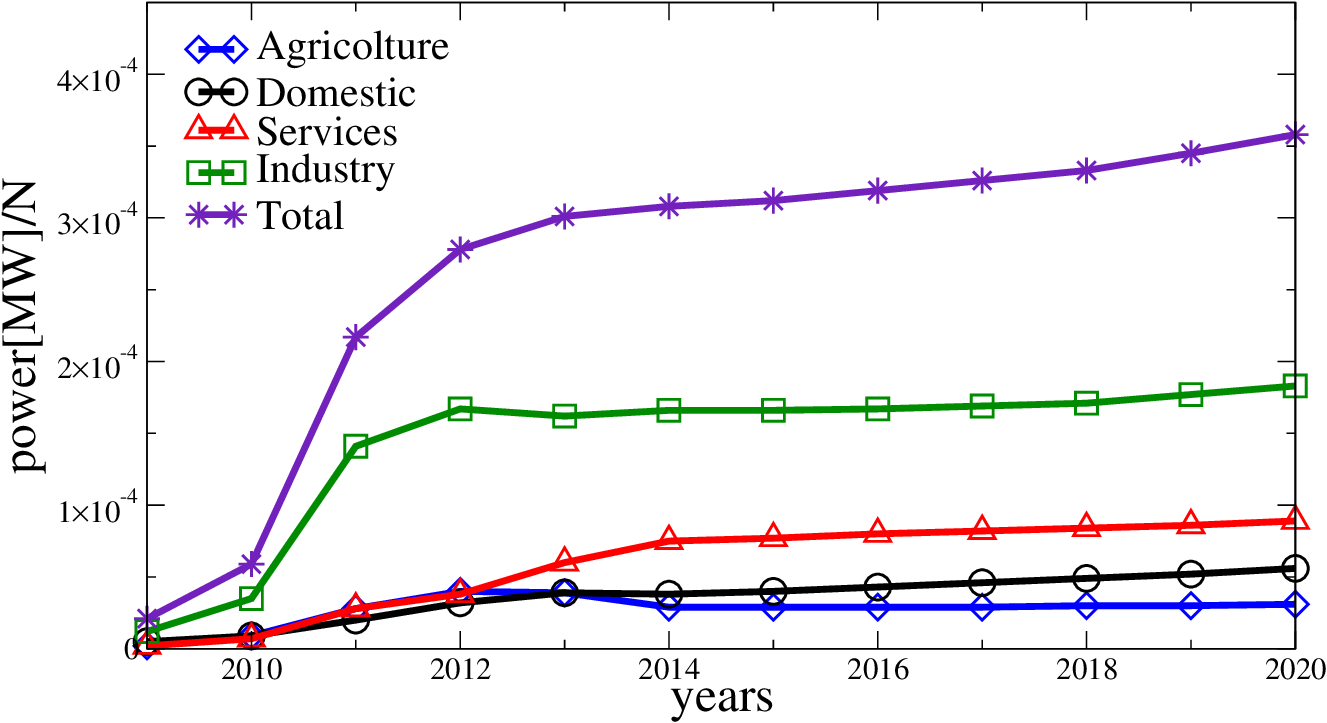}}
\vspace*{8pt}
\caption{Decomposition of Italy time-series in main sectors.}\label{fig:sectors}
\end{figure}

\textcolor{black}{The second important image for our analysis is the breakdown of total power in Italy into four categories, based on data from~\cite{GSE}. These categories are: ``Industry" (which makes up more than half of Italian power), ``Domestic", ``Services", and ``Agriculture". The time-series data are shown in Figure \ref{fig:sectors}. In the following sections, as discussed in the Introduction, we will focus on ``Industry" and ``Domestic".
}

\section{The Markov model}\label{sec:Markov}
\textcolor{black}{This section presents the Markov model capturing the dynamics of the ``Domestic" time-series in Figure \ref{fig:sectors}. We aim to capture the idea that the ``Domestic" compartment answers to incentives in the following way: it just increases \textit{incrementally} its attitude to use them, and it does not enter into a game logic, as done by the ``Industry" compartment.\\
\indent Toward this aim, let $N \in \mathbb{N}$ be the number of individual people in our system, each one characterized by a state $X_t^{i}$, $i \in \{1,\ldots,N\}$, at time $t$ taking one of the following two \textit{qualitative} values: $X_t^{i} \in \{D, G\}$ The state $D$ stands for \textit{Deliberating}, whereas $G$ for \textit{Green}. The former term comprises individual people or householders interested in installing solar panel photovoltaics; in principle, they want to participate but have not done so yet. In particular, we are assuming that such individual people can articulate a complex process lying behind the assumption of a specific (pattern of) action, in this case, the installation of solar photovoltaic panels. They assess specific starting conditions, the appropriateness of the time for acting and the
effective means at their disposal, in addition to an overall and, at the same time, analytical evaluation of the consequences that the adoption of -- or the refusal of adopting -- that specific (pattern of) action may produce. Notice that a complete model of the whole population should include other classes, like those who do not have in mind solar panels or are even against it; e.g., one can think of the so-called \emph{Sinus-Milieus}$^{\tiny{\text{\textregistered}}}$ categories mentioned in \cite{Palmer}. The state $G$, instead, means \textit{Green}, namely that the subject has installed the solar photovoltaic panels. The state $G$ is absorbing: a subject may jump from $D$ to $G$ but cannot jump back from $G$ to $D$.\\
\indent One possible way to describe the jump from the state $D$ to either the state $G$ or $D$ could be the usage of a discrete-time Markov chain with suitable transition probabilities $p_{D \rightarrow D}$ or $p_{D \rightarrow G}$, where $p_{D \rightarrow D} = 1 - p_{D \rightarrow G}$. However, since we will look for equations satisfied in the limit when the number of individual people $N$ tends to infinity, which we will call, with a slightly abuse of language, ``mean-field" equations, we opt for the usage of continuous-time Markov models since they have better analytical rules like the so-called Dynkin formula.  In an attempt to make this paper more self-contained, Section \ref{dynkin and Aldous} in the Appendix reviews some aspects of continuous-time Markov models; \textit{rates of transitions} will replace the transition probabilities.\\
\indent To describe the rates of transitions, let $\mathcal{S}_N:=\{D,G\}^{N}$ be the state-space of the full system and $x_N = (x^1,\ldots, x^N) \in \mathcal{S}_N:=\{D,G\}^{N}$ be a \textit{configuration} of the latter, namely an element of the state space.  We suppose that the rate of transition from $D$ to $G$ of a single individual is a deterministic function $\lambda_N(\cdot, \cdot)$ of an element of the full state space and time $t$, because we shall include in the model the possibility that the transition rules change in time, for instance as a consequence of new governmental policy about solar panels. We select the following straightforward but specific form for the rate of transition:
\begin{equation}\label{eq::transitionratemarkov}
   \lambda_{N}\left(  x_N,t\right)  =a\left(  t\right)  \frac{N_{G}\left(x_N\right)}{N},\,\,\text{where}\,\,N_{G}\left(x_N\right)  =\sum_{j=1}^{N}1_{\left\{  x^{j}=G\right\}}
\end{equation}
is the number of green individual people at time $t$; in particular, our Markov chain is a non-homogeneous Markov chain (NHMC). We now explain the intuition behind the previous choice.   The Markov nature, without any decision/optimization aspect, reflects, in
the simplest possible modelling way, the idea of bounded rationality; as said in the introduction, for the sake of presentation only, in what follows, we will speak about procrastination. Each subject, at every time instant $t$, has a rate to install solar photovoltaic panels instead of procrastinating. The
dependence of $\lambda_{N}\left(  x_N,t\right)  $ from $\frac{N_{G}\left(x_N\right)  }{N}$ corresponds either to the idea that we have a tendency to
imitate the others, or to the more impersonal fact that, when a larger number
of Greens exist, the Market is more developed -- maybe lower prices, better
distribution network, more information and experience -- and the probability of installation increases. Instead, the factor $a(t)$ modulates the rate to give importance to external factors like subsidies. We will see in Subsection \ref{subsec::simulationmarkov} that $a(t)$ will be modified over the period under scrutiny in a way that is proportional to the number of incentives, to capture the ``Domestic" attitude to use incentives described at the beginning of the present section. 
\begin{remark}
At an abstract level, we can incorporate additional elements into the model, such as personal characteristics of the individual that may influence the decision to install panels. This could include an additive term structured like \(\lambda_{N}\left( x_N,t \right) = a\left(t\right) \frac{N_{G}\left( x_N \right)}{N} + b\left(t\right)\) in the rate, or a stochastic process representing external inputs, which would reflect a dynamic rather than simply being captured by the time-dependence of \(a\left(t\right)\) and \(b\left(t\right)\). Another idea for enhancing the model's realism and introducing intriguing mathematical complexities would be to allow the characteristics of individuals—specifically, whether they are players or not—to evolve over time. This could be modeled using a jump process that takes values in \(\{0,1\}\) (see \cite{blume1993statistical, sandholm2010population}). However, we opted for the simpler model, as it is sufficient to successfully fit the data.
\end{remark}
\indent Now, we define the infinitesimal generator of the continuous-time Markov process introduced above.    To this end, let $F$\footnote{In order to avoid burdening the notation, we omit the dependence of $F$ by the number of individual people $N$ in the present section.} be a real-valued test function on $\mathcal{S}_N$ which therefore depends only on a finite number $N$ of coordinates, $F = F(x_N)=F(x^{1},\ldots,x^{N})$ with $x^{i} \in \{D,G\}$. The time-dependent infinitesimal generator is given by:
\begin{equation}\label{eq::infinitesimalgenerator}
\left(\mathcal{L}_{t,N}\right)\left( x_N\right)     =\sum_{i=1}^{N}1_{\left\{
x^{i}=D\right\}  }\lambda_{N}\left(x_N,t\right)  \left(  F\left(
x^{i\rightarrow G}_N\right)  -F\left(  x_N\right)  \right),
\end{equation}
where $\lambda_{N}\left(x_N,t\right)$ has been defined in Equation \eqref{eq::transitionratemarkov} and $x^{i \rightarrow G}_N$ is the configuration $x_N$ where we have imposed $x^{i} = G$. The intuition behind the generator in Equation \eqref{eq::infinitesimalgenerator} is that each subject $i$ which is Deliberating (the sum is restricted to Deliberating ones by the factor $1_{\left\{x^{i}=D\right\}  }$) has a rate $\lambda_{N}\left(x_N,t\right)  $ to become Green. Heuristically, we could write in the limit as $\Delta
t\rightarrow0$:%
\[
\text{Prob}\left(  X_{t+\Delta t}= G |X_{t}=G, x_N\right)  \sim\lambda_{N}\left(
x_N,t\right)  \cdot\Delta t,
\]
where \text{Prob} denotes the probability.  Now, in the next subsection we derive the Dynkin equations, whereas in Subsection \ref{subsec::limitmodel} the limit model.  
\subsection{Dynkin equations}\label{subsec::dynkinequation}
\indent One of the advantages of the continuous-time modelling is the possibility to write an identity, similar to an integral equation, for the percentage of green subject at time $t$, i.e.
\begin{equation}\label{eq::percentageofgreen}
    p_G^N(t):=\frac{N_G(X_t^{N})}{N},
\end{equation}
where $X_t^{N}:=(X_t^1,\ldots,X_t^N)$; we are aware of a clash in the notation but it will be clear from the context when referring to the last entry of $X_t^{N}$. This is the content of the next proposition.
\begin{proposition}\label{prop::markov green}
Let $N \in \mathbb{N}$, and assume that $(X_t^{N})_{t \geq 0}$ is the continuous-time Markov process with time-dependent infinitesimal generator given in Equation \eqref{eq::infinitesimalgenerator}. Then, the percentage of green subject $p_G^{N}(t)$ at time $t$ in Equation \eqref{eq::percentageofgreen} satisfies the following equation:
\begin{equation}\label{Ito Dynkin}
    p_{G}^{N}\left(  t\right)  =p_{G}^{N}\left(  0\right)  +\int_{0}^{t}a\left(
s\right)  \left(  1-p_{G}^{N}\left(  s\right)  \right)  p_{G}^{N}\left(
s\right)  ds+M_{t}/N,
\end{equation}
where $(M_t)_{t \geq 0}$ is a zero-mean martingale with a variance bounded above by $N \int_{0}^{t} a(s)\,ds$.
\end{proposition}
\begin{proof}
Let $\mathcal{L}_t F(x_N)$ be the infinitesimal generator in Equation \eqref{eq::infinitesimalgenerator} when we consider as $F$ the test function $F(x_N)=N_G(x_N)$, with $N_G(x_N)$ defined in Equation \eqref{eq::transitionratemarkov}. We have:
\begin{equation*}
\begin{split}
    \mathcal{L}_{t,N} N_G(x_N) &= \sum_{i=1}^{N} 1_{\left\{
x^{i}=D\right\}  }\lambda_{N}\left(x_N,t\right)\left(N_G(x^{i \rightarrow G}_N)-N_G(x_N)\right)\\
                       &= \sum_{i=1}^{N} 1_{\left\{x^{i}=D\right\}} \lambda_N(x_N,t) = a(t) \frac{(N-N_G(x_N))N_G(x_N)}{N},
\end{split}
\end{equation*}
where we have used the fact that, for each configuration $x_N \in \mathcal{S}_N$ and each $i \in \{1,\ldots,N\}$, we have that $1_{\left\{x^{i}=D\right\}} \left(N_G(x^{i \rightarrow G}_N)-N_G(x_N)\right) = 1_{\left\{x^{i}=D\right\}}$. By the first Dynkin's formula (see the Appendix, Section \ref{dynkin and Aldous}, Equation \eqref{eq::firstandsecond}) we have that:
\begin{equation*}
N_{G}\left(  X_t^N  \right)  =N_{G}\left(  X_0^N\right)
 +\int_{0}^{t}a\left(  s\right)  \frac{\left(  N-N_{G}\left(  X_s^N  \right)  \right)  N_{G}\left(  X_s^N  \right)  }%
{N}ds+M_{t}%
\end{equation*}
where $M_{t}$ is a martingale; we have omitted the explicit dependence upon the test function. Equation \eqref{Ito Dynkin} follows by dividing both sides of the previous equation by $N$.\\
\noindent   In order to estimate the variance of the martingale $M_t$, we make use of the second Dynkin's formula (see, again, the Appendix, Section \ref{dynkin and Aldous}, Equation \eqref{eq::firstandsecond}). In particular, we need to compute $(\mathcal{L}_{t,N}F_N^2-2 F_N \mathcal{L}_{t,N}F_N)(x_N)$ with $F_N(x_N) = N_G(x_N)$. We have
\begin{equation*}
    \begin{split}
        (\mathcal{L}_{t,N}F_N^2-2 F_N \mathcal{L}_{t,N}F_N)(x_N)  &=\sum_{i=1}^{N} 1_{\{x^{i}=D\}}\lambda_N(x_N,t)(N_G(x^{i \rightarrow G})-N_G(x_N))^2\\
        &=\sum_{i=1}^{N} 1_{\{x^{i}=D\}}\lambda_N(x_N,t) = a(t)\frac{(N-N_G(x_N))N_G(x_N)}{N}.                                                   
    \end{split}
\end{equation*}
Therefore, $|(\mathcal{L}_{t,N}F_N^2-2 F_N \mathcal{L}_{t,N}F_N)(x)|\leq a(t)N$. Thus:
\begin{equation*}
    \mathbb{E}[M_t^2] = \mathbb{E}\left[\int_{0}^{t}(\mathcal{L}_s F_N^2-2 F_N \mathcal{L}_s F_N)(X_s^N)\,ds\,\right] \leq N \int_{0}^{t} a(s)\,ds.
\end{equation*}
This concludes the proof of the proposition.
\end{proof}
Equation \eqref{Ito Dynkin} states that the change $p_G^N(t)-p_G^N(0)$ in a time interval $[0,T]$,  of the percentage of green individual people is given by the time-integral of a logistic term, $(1-p_G^N(s))p_G^N(s)$, modulated by the factor $a(t)$, plus a reminder $M_t/N$, where $M_t$ is a martingale.}\\
\indent Finally, in the next subsection, we describe the limit system for the $N$-subject system introduced above.
\subsection{Limit model}\label{subsec::limitmodel}
\textcolor{black}{We now prove the convergence of the stochastic process $(p_G^N(t))_{t \geq 0}$ to the unique solution of a deterministic Cauchy problem.    Notice that for most scaling limit problems, this target is achieved by a series of steps:
\begin{itemize}
    \item[(i)]  First one proves a tightness criteria for $p_G^N(t)$; this easily follows from Proposition \ref{prop::tightness} in Appendix, Section \ref{dynkin and Aldous}.
    \item[(ii)] Then, by Prohorov's theorem, there exist subsequences of the family $(P_N)_{N \in \mathbb{N}}$ of laws of $p_G^N(\cdot)$ which converge in the sense of weak convergence of measures on a suitable path space; in our case the path space is the Skorohod space $\mathcal{D}(0, T)$.
    \item[(iii)]    Via classical arguments -- one of them based on Skorohod representation theorem, the other one based on the method illustrated in \cite{KL}, Page 56 -- one proves that the limit points of $(P_N)_{N \in \mathbb{N}}$ are concentrated on solutions of the deterministic Cauchy problem. 
    \item[(iv)]     By a classical result of uniqueness for the deterministic Cauchy problem, all limit points of $(P_N)_{N \in \mathbb{N}}$ are the same delta Dirac measures $\delta_p$ at the above mentioned unique solution.
    \item[(v)]      Therefore, the entire sequence $(P_N)_{N \in \mathbb{N}}$ convergence to $\delta_p$; this follows from the fact that the weak convergence is a metric convergence. 
    \item[(vi)]     Finally, the weak convergence is upgraded to a convergence in probability since the weak limit point is deterministic.
\end{itemize}
\noindent   It is important to notice that the strategy above works well in our case, but it provides only a convergence in probability. In the next proposition, we will manage to prove that $p_G^N(\cdot)$ converges locally uniformly in $t$, in a mean square (whence in probability) to $p_G(t)$ with a special trick, which is, however, not a universal one.
\begin{proposition}\label{Limit meanfield}
    Let $N \in \mathbb{N}$, and $p_G(t)$ be the unique solution of the following Cauchy problem
    \begin{equation}\label{eq::ode}
        \begin{split}
            &\frac{d p_G(t)}{dt} = a(t) (1-p_G(t)) p_G(t),\\
            &p_G(0) = p_G^0, 
        \end{split}
    \end{equation}
for a given initial condition $p_G^0 \in [0,1]$. Moreover, assume that the sequence of random variables $(p_G^{N}(0))_{N \in \mathbb{N}}$ converges to $p_G^0$ in mean square; let $c_N$ be the following quantity
\begin{equation*}
    c_N:=\mathbb{E}[\left|p_G^N(0)-p_G^0\right|^2].
\end{equation*}
\end{proposition}
Then, $p_G^N(\cdot)$ converges locally uniformly in $t$, in a mean square (whence in probability) to $p_G(t)$. Precisely, we have
\begin{equation*}
    \mathbb{E}\left[\sup_{t \in [0,T]}|p_G^N(t)-p_G(t)|^2\right] \leq 2 e^{4 A(T)}\left(c_N + \frac{4 A(T)}{N}\right),
\end{equation*}
where $A(t)=\int_{0}^{t}a(s)\,ds$.
\begin{proof}
First, we rewrite the differential equation in \ref{eq::ode}  in integral form. We have:
\begin{equation*}
    p_G(t) = p_G^{0} + \int_{0}^{t} a(s)(1-p_G(s))p_G(s)\,ds.
\end{equation*}
Then, we compare the previous equation with the Dynkin's identity in Equation \eqref{Ito Dynkin} by estimating the following difference:
\begin{equation*}
    \begin{split}
        q_N(t)  &:=\left|p_G^N(t)-p_G(t)\right|\\
                &\leq q_N(0) + |M_t|/N + \int_0^t a(s)\left|(1-p_G^N(s))p_G^N(s)-(1-p_G(s))p_G(s)\right|\,ds\\
                &\leq q_N(0) +  |M_t|/N\\
                &+\int_0^t a(s)\left|(p_G(s)-p_G^N(s))p_G^N(s)\right|\,ds\\
                &+\int_0^t a(s)\left|(1-p_G(s))(p_G^N(s)-p_G(s))\right|\,ds\\
                &\leq q_N(0) +  |M_t|/N\\
                &+\int_0^t a(s)q_N(s)p_G^N(s)\,ds + \int_0^ta(s)(1-p_G(s))q_N(s)\,ds\\
                &\leq q_N(0) + |M_t|/N + \int_{0}^{t} 2 a(s) q_N(s) \, ds.
    \end{split}
\end{equation*}
By the Gronwall's lemma we have 
\begin{equation*}
    q_N(t) \leq e^{\int_{0}^{t}2 a(s)\,ds} (q_N(0) + |M_t|/N).
\end{equation*}
In particular, 
\begin{equation*}
    \mathbb{E}\left[\sup_{t \in [0,T]} q^n_N(t)\right] \leq 2 e^{\int_0^T 4 a(s)\,ds}\left(\mathbb{E}\left[\left|p_G^0-p_G(0)\right|^2\right]+\mathbb{E}[|M_t|^2]/N^2\right).
\end{equation*}
The claims of the proposition follow from the estimate on $\mathbb{E}[|M_t|^2]$ as in Proposition \ref{prop::markov green} together with the Doob's inequality for martingales.
\end{proof}}

\subsection{Simulations}\label{subsec::simulationmarkov}
\begin{figure}[!h]
\centerline{\includegraphics[width=3.8in]{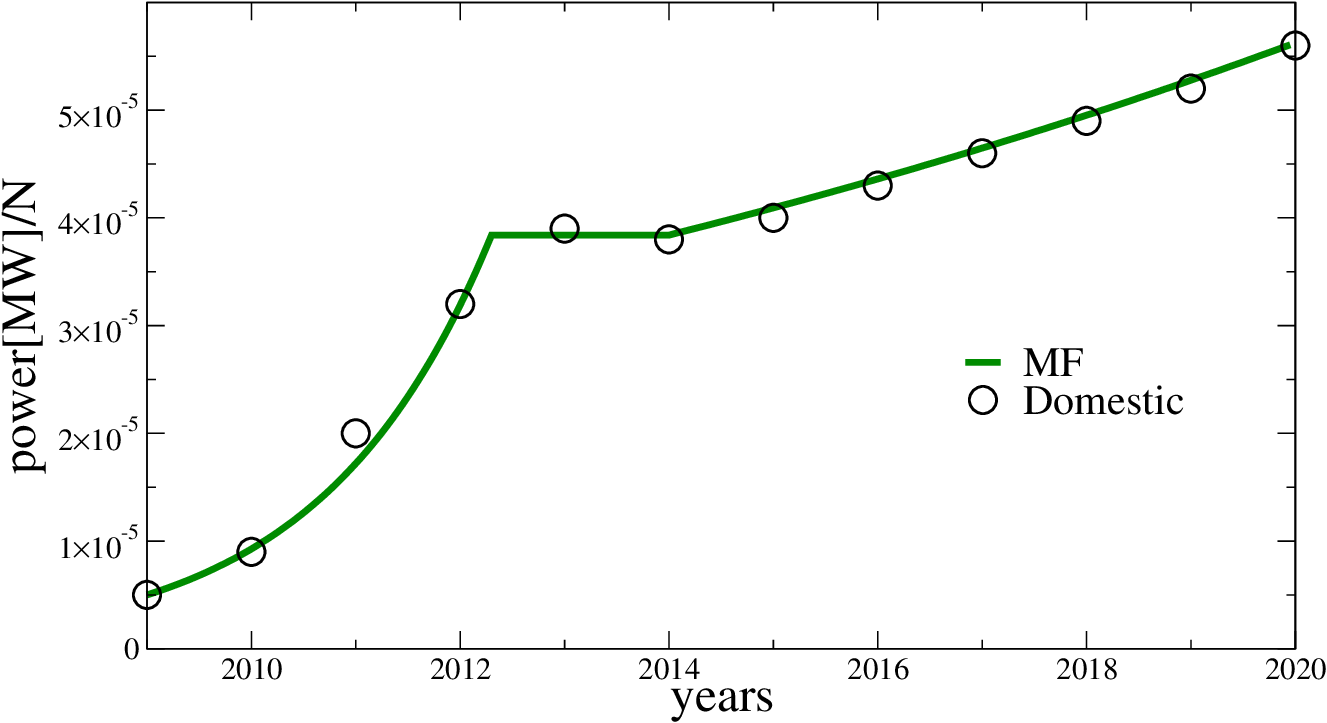}}
\vspace*{8pt}
\caption{Mean-Field model (MF) is the continuous curve in green, over the data (small circles).}\label{fig:mf}
\vspace*{16pt}
\centerline{\includegraphics[width=3.8in]{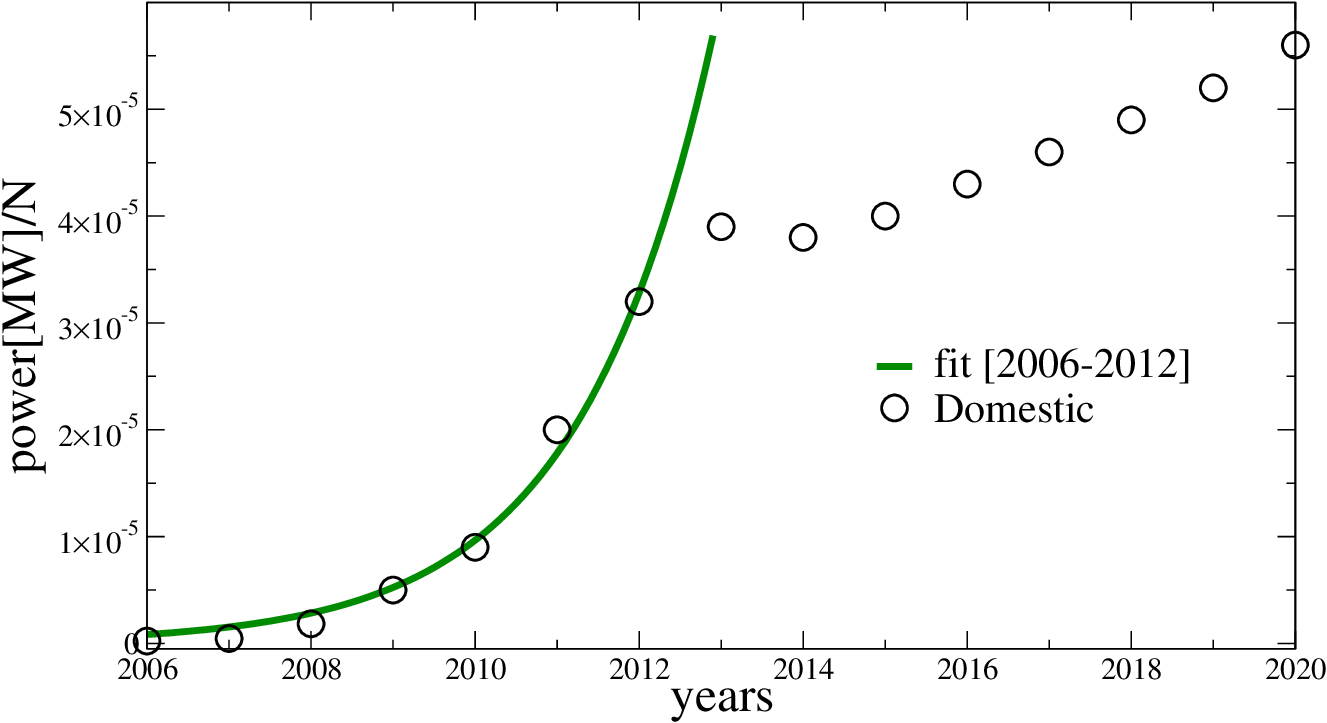}}
\vspace*{8pt}
\caption{The fit of the first period, based on extrapolated initial data, and its potential continuation.}\label{fig:exp}
\end{figure}

As described in the \textcolor{black}{section} Introduction, we apply the Markov model to the ``Domestic" data, since it is natural to conjecture that procrastination, which is at the foundation of the choice to use a Markov chain, is proper of \textcolor{black}{individual people} or householders (instead of companies, as discussed below in \textcolor{black}{Section \ref{sec:Meanfieldgameapproach}.)}\\
\indent The Domestic time-series requires a specification of the way we have obtained
the data, from~\cite{GSE}. In the period 2010-2020 we have explicit
values of the Domestic data; the total national data is divided in the four
categories described in \textcolor{black}{Section \ref{sect time-series}. In 2009 there is a different subdivision in categories from which we deduce that Domestic is approximately
given by a certain percentage $p_{2009}$ of the total; we have used this datum
in the plot. Concerning 2006-2008 we do not even have the percentage.\\
\indent We have used the \textcolor{black}{limit system}, i.e. the mean-field, derived in the previous section, namely the ordinary differential equation for the quantity $p_{G}\left(  t\right)  $, with the following values:
\begin{align*}
a\left(  t\right)   &  =0.627\text{ for }t\leq 2012\\
a\left(  t\right)   &  =0.0637\text{ for }t\geq2014
\end{align*}
and a constant connection in between, in absence of a better understanding of
the intermediate period between the two different subsidy regimes. The result
is shown in Figure \ref{fig:mf}. The values of $a\left(  t\right)  $ have been fitted, by an exponential fit with minimum mean square error, considering the two separate time periods; then they have been used in the differential equation model, with suitable
initial condition. Since the fit in the period 2009-2012 may look arbitrary due to the smallness of the time series (only 4 values), we extrapolated values of the Domestic
time series by using the proportion coefficient $p_{2009}$ also in the years
2006-2008. The fit of the first period is surprisingly the same (first three
digits) hence it looks particularly stable. The result is shown in Figure \ref{fig:exp}.%
If we rely on this stable fit, we may guess that the increase of Domestic in
the next few years would lead to an excellent performance, as the green
extrapolation line of Figure \ref{fig:exp} shows. Unfortunately, it was decided to change the structure of the subsidies.}

\section{The Mean-Field Game approach}\label{sec:Meanfieldgameapproach}
\textcolor{black}{This section presents the mean-field game system capturing the dynamics of the ``Industry" time-series in Figure \ref{fig:sectors}.    The mean-field game we are going to present is a continuous-time finite state mean-field game. We will use the narrative and the formalism used in \cite{gomes2013continuous}. In particular, we first describe, in Subsection \ref{sec::meanfield}, a mean-field model which corresponds to the limit as the number of players -- companies or firms will be called players in this context -- tends to infinity of a symmetric dynamic games with a finite number of players, which will be presented in Subsection \ref{sec::nplayer}.  Instead, we refer to \cite{gomes2013continuous}, Theorem 7 in Section 4, for the proof of the convergence, as the number of players $N\rightarrow\infty$ in $L^2$ of the $N+1-$player model to the mean field model of Subsection \ref{sec::meanfield}. Admittedly, a more realistic setting would be to add to our model common noise events at fixed points in time as in \cite{belak2021continuous}, which would lead to a system of random forward-backward ODEs foe the mean field equilibrium. However, this would be an additional technicality that would not add to the present work's conceptual advancements.
\subsection{The mean-field model}\label{sec::meanfield}
We begin by presenting a mean-field model for a continuous-time dynamic game between many rational agents, represented by firms or companies, which, as said, we call players. These players are allowed to switch to the state $G$, which stands for \textit{Green}, if they are in the state $D$, which stands for \textit{Deliberating}. We assume that they look forward to optimizing a specific functional, which depends on the statistical distribution of the other players. The state $G$, instead, is absorbing. We suppose that all players are identical, so the game is symmetric concerning the permutation of the players. Players in a particular state only know their state and the fraction of players in each of the two considered states. In particular, each player in state $D$ can control the transition rate from $D$ to $G$ and incurs a running cost which depends on its own state, on the state of the other players (through its distribution among states but not on individual player's states), as well as on the control the player chooses.\\
\indent We now fix one of the players which will be called the reference player; because the game is symmetric, the player's identity is not important, and all other players have access to similar information. As said, we further assume the mean-field hypothesis, that is, since the number of players is huge, the only information available to the reference player is the distribution of the other players. In particular, we suppose that the players' distribution among states is given by a probability vector $\theta(t)=(p_G(t), p_D(t)) \in \mathcal{I}^{2}$, where $\mathcal{I}^2$ is the probability simplex
\begin{equation*}
    \begin{cases}
        p_G(t)+p_D(t)=1,\\
        p_G(t) \geq 0,\,\,p_{D}(t) \geq 0.
    \end{cases}
\end{equation*}
Let $\Lambda(t) \in \mathbb{R}^{2 \times 2}$ represent a transition rate matrix depending on the time $t$, where $\Lambda_{D \rightarrow G}(t) \geq 0$, $\Lambda_{G \rightarrow D}(t) \geq 0$, $\Lambda_{G \rightarrow G}(t)=-\Lambda_{G \rightarrow D}(t)$ and $\Lambda_{D \rightarrow D}(t)=-\Lambda_{D \rightarrow G}(t)$. We assume that the players switch from state to state according to a continuous-time (non homogeneous) Markov process with transition rate matrix $\Lambda$. In the mean-field limit, the fraction of players in each state $\theta$ satisfies a Kolmogorov equation, which is complemented by an initial condition from which the evolution of the distribution of players $\theta^{\Lambda}:[0,T] \rightarrow \mathcal{I}^2$ is completely determined.
\subsubsection{Running costs, and single player control problem: the value function}
\indent We fix now a reference player and consider the running cost in an optimization problem according to its point of view. In general, the running cost is a function $c : \{D, G\} \times \mathcal{I}^2 \times (\mathbb{R}_0^{+})^2 \rightarrow \mathbb{R}$, $c(s_i, \theta, \lambda)$ where $\mathbb{R}_0^{+}$ is the set of real numbers greater or equal than zero, $\theta \in \mathcal{I}^2$ is the probability distribution of players among states, and $\lambda_{s_i \rightarrow s_j}$ is the transition rate the reference player uses to change from state $s_i$ to state $s_j$. In particular, we are interested in the running cost of the reference player whose state at the initial time is $D$; if the state at the initial time is $G$, instead, the reference player does not incur any cost. Suppose the players are distributed among the state $D$ and $G$ according to the distribution probability $\theta:[0,T] \rightarrow \mathcal{I}^2$, which for now we assume to be known by the reference player. Let 
\begin{equation}\label{eq::costfunctional}
    v^{D}_{\theta}(t, \lambda) = \mathbb{E}^{\lambda}_{X_t = D}\left[\int_{t}^{T}c(X_s, \theta(s), \lambda(s))\,ds\right] = \mathbb{E}^{\lambda}_{X_t = D}\left[\int_{t}^{T}\left(h(\lambda_{D \rightarrow G}(s)) + p_G(s)\right)1_{\{X_s=D\}}\,ds\right],
\end{equation}
where $\mathbb{E}^{\lambda}_{X_t = D}$ is the expectation conditioned on the event $\{X_t=D\}$, given the transition rate $\lambda$. We now comment on the meaning of the functional in Equation \eqref{eq::costfunctional}. As regards the first part of the running cost $h(\lambda_{D \rightarrow G}(s))1_{\{X_s=D\}}$, we suppose that $h(\cdot) \geq 0$ is an increasing function with $h(0) = 0$. We interpret it as being symptomatic of the fact that choosing a transition rate $\lambda_{D \rightarrow G}(s) \geq 0$ has a psychological cost $h(\lambda_{D \rightarrow G}(s))$; this produces a tendency to procrastinate. Notice that the representative player pays this cost until it makes the transition to $G$. Once it is in the state $G$, the just-mentioned psychological cost is no longer present; admittedly, in this way, we are neglecting other types of costs that the representative agent can incur once that it makes the transition; e.g. maintenance costs of solar panels. Instead, the term $\mathbb{E}^{\lambda}_{X_t = D}\left[\int_{t}^{T} p_G(s) 1_{\{X_s=D}\}\,ds\right]$ can be interpreted in the following way. Assume there is a limited amount of resources for a certain subsidy. If $p_{G}(s)$ is high, meaning that several subjects have already utilized the subsidy, it becomes more challenging to obtain it, resulting in a higher cost. On the other hand, $v_{\theta}^{G}(t,\lambda)=0$. \\
\indent We now define the value functions associated to $\theta$, denoted by $v_{\theta}:[0,T] \times \{D,G\} \rightarrow \mathbb{R}$, namely the costs incurred by the representative player corresponding to the two initial conditions $(t,D)$ and $(t,G)$. In particular, we will study only the function $v_{\theta}^{D}$. We have that:
\begin{equation}\label{eq::value function}
    \begin{split}
        v_{\theta}^{D}(t)=\min_{\lambda} v_{\theta}^{D}(t,\lambda),
    \end{split}
\end{equation}
where the minimization is performed over Markovian controls $\lambda(s) = \lambda(X_s,s)$, with $(X_s)_{s \geq 0}$ a continuous-time Markov chain controlled by $\lambda$ which corresponds to the state of the reference player at time $s$. More precisely $\text{Prob}(X_{s+\Delta t}=G|X_{s}=D) \sim \lambda_{D \rightarrow G}(X_s,s) \cdot \Delta t$.\\ 
\indent By continuing to assume that $\theta$ is given, as in the classical optimal control, we introduce now the Hamilton-Jacobi-Bellman ODE for $v_{\theta}^{D}$. In general, the latter equation written for a value function $v_{\theta}(t, x)$, where $t \in [0,T]$ and $x \in \{D,G\}$ is given by:
\begin{equation}\label{eq::value function D}
    \min_{\lambda \geq 0} \{\partial_t v_{\theta}(t,x) + \mathcal{L}^{\lambda}_t v_{\theta}(t,x) + (h(\lambda(t)) + p_G(t))1_{\{x = D\}}\}=0,
\end{equation}
where $\mathcal{L}^{\lambda}_t$ is the $\lambda$-dependent infinitesimal generator of the continuous-time Markov process $(X_s)_{s \geq 0}$ with state-space $\{D,G\}$. Moreover, $v_{\theta}(T)=0$. In the particular case of the function $v_{\theta}^{D}$ we have that the previous equation is given by
\begin{equation}\label{eq::hjb for vD}
    \min_{\lambda_{D \rightarrow G} \geq 0} \{\partial_t v_{\theta}^{D}(t) + \mathcal{L}^{\lambda}_t v_{\theta}^{D}(t) + (h(\lambda_{D \rightarrow G}(t)) + p_G(t))\}=0,
\end{equation}
with $v_{\theta}^{D}(T)=0$; in particular, we need to solve a terminal value problem. We now specialize the previous Equation \eqref{eq::hjb for vD} in the simple choice that $h(\lambda) = \sigma^2 \lambda^2$, where $\sigma^2>0$ is a positive constant. Considering a more general form for the cost $h(\lambda)$ would be an additional technicality that would not add to the present work's conceptual advancements. The infinitesimal generator of the process $(X_s)_{s \geq 0}$ associated to the value function $v_{\theta}$ is given by
    \begin{equation*}
        \mathcal{L}_{t}^{\lambda} v_{\theta}(t,x) = 1_{\{x=D\}}\lambda_{D \rightarrow G}(t)\left(v_{\theta}(t,G)-v_{\theta}(t,x)\right),
    \end{equation*}
which means that if the state of representative player is $G$, then it remains $G$; if it is $D$, then the rate of transition to $G$ is $\lambda_{D \rightarrow G}(t)$.   In particular, we have that $\mathcal{L}_{t}^{\lambda} v_{\theta}^{D}(t) = - \lambda_{D \rightarrow G}(t) v_{\theta}^{D}(t)$. Therefore, the Hamilton-Jacobi-Bellmann terminal value problem in Equation \eqref{eq::value function D} becomes
\begin{equation*}
    \partial_t v_{\theta}^{D}(t) + p_G(t) + \min_{\lambda_{D \rightarrow G}(t) \geq 0}\{h(\lambda_{D \rightarrow G}(t))-\lambda_{D \rightarrow G}(t) v_{\theta}^{D}(t)\} = 0.
\end{equation*}
In our case $h(\lambda) = \sigma^2 \lambda^2$. Then, the minimum of the function $\lambda \rightarrow \sigma^2 \lambda^2 - \lambda v_{\theta}^{D}(t) = \lambda (\sigma^2 \lambda - v_{\theta}^{D}(t))$ is the solution of $2 \sigma^2 \lambda = v_{\theta}^{D}(t)$. Whence, we obtain:
\begin{equation*}
    \lambda^{*}_{D \rightarrow G}(X_t,t) = \lambda^{*}_{D \rightarrow G}(v_{\theta}^{D}, \theta(t), D) = \frac{1}{2 \sigma^2} v_{\theta}^{D}(t),
\end{equation*}
which will be the optimal control. The minimum value of the function $\lambda \rightarrow \sigma^2 \lambda^2 - \lambda v_{\theta}^{D}(t)$ in correspondence of such control is equal to $-\frac{1}{4 \sigma^2} v_{\theta}^{D}(t)$. In particular, the Hamilton-Jacobi-Bellmann ODE equation for $v_{\theta}^{D}(t)$ is given by
\begin{equation*}
    \begin{cases}
    \partial_t v_{\theta}^{D}(t) + p_G(t) - \frac{1}{4 \sigma^2} v_{\theta}^{D}(t) = 0\\
    v_{\theta}^{D}(T)=0.
    \end{cases}
\end{equation*}
We have just proved the following Theorem.
\begin{theorem}\label{th::verification}
    Suppose $v_{\theta}^{D}:[0,T]\rightarrow\mathbb{R}$ is a solution on $[0,T]$ of the following Hamilton-Jacobi-Bellmann ODE : 
    \begin{equation*}
    \begin{cases}
    \partial_t v_{\theta}^{D}(t) + p_G(t) - \frac{1}{4 \sigma^2} v_{\theta}^{D}(t) = 0\\
    v_{\theta}^{D}(T)=0.
    \end{cases}
\end{equation*}
Then $\lambda^{*}_{D \rightarrow G}(X_t,t) = \lambda^{*}_{D \rightarrow G}(v_{\theta}^{D}, \theta(t), D) = \frac{1}{2 \sigma^2} v_{\theta}^{D}(t)$ is the optimal Markovian transition rate from $D$ to $G$.
\end{theorem}
\subsubsection{Mean-field Nash equilibrium}
The mean field Nash equilibrium occurs when the background players are using a strategy $\Lambda$ for which the best response of the reference player is $\Lambda$ itself. It means that each background player, say $X^1$ wants to solve the same problem in the previous subsection but where $p_{G}(t)$ is replaced by the probability $\widetilde{p}_G(t)$ that the player $X^1$ is in the state $G$ at time $t$, when the optimal strategy is chosen. Hence, the dynamics of the subject $X^1$ is given by the following time-dependent infinitesimal generator:
\begin{equation*}
    \widetilde{\mathcal{L}}_t F(x^1) = 1_{\{x^1=D\}}\frac{1}{2\sigma^2}\widetilde{v}_D(t)(F(x^{1\rightarrow G})-F(x^1)),
\end{equation*}
where the function $\widetilde{v}_D(t)$ solves Hamilton-Jacobi-
Bellmann ODE in Theorem \ref{th::verification} associated to $\widetilde{p}_G(t)$, and $F$ is a test-function. We now apply the first Dynkin's formula (see Appendix, Section \ref{dynkin and Aldous}, Equation \eqref{eq::firstandsecond}) to the test function $F(x^1)=1_{\{x^1=G\}}$. We have that:
\begin{equation*}
    1_{\{X_t^1=G\}} = 1_{\{X_0^1=G\}} + \int_{0}^{t}\widetilde{\mathcal{L}}_s F(X_s)\,ds + M_t,
\end{equation*}
where $M_t$ is a martingale.    Because $\widetilde{\mathcal{L}}_t F(x^1) = 1_{\{x^1=D\}}\frac{1}{2 \sigma^2}\widetilde{v}_D(t)$, we have that:
\begin{equation*}
    1_{\{X_t^1=G\}} = 1_{\{X_0^1=G\}} + \int_{0}^{t} 1_{\{X_s^1=D\}}\frac{1}{2 \sigma^2}\widetilde{v}_D(s)\,ds + M_t.
\end{equation*}
By taking the expected value on both sides, we obtain:
\begin{equation*}
    \widetilde{p}_G(t)=\widetilde{p}_G(0)+\int_0^t(1-\widetilde{p}_G(s))\frac{1}{2 \sigma^2}\widetilde{v}_D(s)\,ds.
\end{equation*}
We have just proved the following Theorem
\begin{theorem}\label{th:nash}
    The mean-field Nash equilibrium is characterized by the following system of Kolmogorov and Hamilton-Jacobi-Bellman ODE equations:
    \begin{equation*}
        \begin{cases}
            \partial_t \widetilde{v}_D(t) + \widetilde{p}_G(t) - \frac{1}{4 \sigma^2}\widetilde{v}_D^2(t)=0\\
            \frac{d}{dt}\widetilde{p}_G(t) = (1-\widetilde{p}_G(t))\frac{1}{2\sigma^2}\widetilde{v}_D(t)
        \end{cases}
    \end{equation*}
    together with the initial-terminal conditions $\widetilde{p}_G(0)=p_0 \in [0,1]$ and $\widetilde{v}_D(T)=0$.
\end{theorem}}
\textcolor{black}{
\subsection{$N$-player game}\label{sec::nplayer}
This subsection consider a game between $N+1$-players which is symmetric under permutation of players. As in the previous subsection, we assume that each player can be in one of the two states $D$ or $G$, and knows, in addition to its state, the number of players in each of the states. We remind that $D$ stands for \textit{Deliberating}, and $G$ stands for \textit{Green}, and we assume that each player in state $D$ controls the transition rate from $D$ to $G$ by incurring a running cost which depends on its own state, on the state of the other players, as well as on the control the player chooses. Players follow a Markovian dynamics, the state $G$ is absorbing, and all players are identical, so the game is symmetric concerning the permutation of the players.  In particular, we adopt the point of view of a reference player, which could be chosen as any of the players. We choose the first one.\\
\indent Let $\mathcal{S}=\{D,G\}$ and 
 $\mathcal{N}_N^{2}=\{(n^{D},n^{G}) \in \mathbb{Z}^2\,|\,n^{D}+n^{G}=N,\,n^{D}\geq 0,\,n^{G}\geq 0\}$. The $N$-player game is characterized by two controlled Markov chains. The first one is the state of the reference player, say the first player, whereas the second one, taking values in $\mathcal{N}_N^2$, records the number of the remaining players (distinct from the reference player) that are in any of the two states at any given time. Each player knows its own state, as well as the number of remaining players that are in any of the two states. No further information is available to any individual player.\\
\indent We analyze the interesting case in which the reference player is in state $D$ and it switches from the state $D$ to the state $G$ according to a switching Markov rate $\lambda_{D\rightarrow G}(n,t)$ which it would like to optimize upon. Besides, for the purpose of the present paper, we explicitly describe the transition probabilities of the state of the players, and we refer to \cite{gomes2013continuous}, Section 3.1, for the ones relates to the second controlled Markov chain.  Let $N_t$ be $N_t = (N_G(X_t^N), N-N_G(X_t^N)) \in \mathcal{N}^2_N$, where $X_t^N = (X_t^1,\ldots,X_t^N)$. We suppose
\begin{equation*}
    \text{Prob}(X_{t+\Delta t}^{1}=G | N_t = n, X_{t}^{1}=D) \sim \lambda_{D\rightarrow G}^{1}(n,t) \cdot \Delta t,
\end{equation*}
and that the the players distinct from the reference player that are in the state $D$ follow controlled Markov process with transition rates from state $D$ to state $G$ given by $\Lambda_{D \rightarrow G}(n,t)$. We suppose that $\lambda^1_{G \rightarrow D}(n,t)=-\lambda^1_{D \rightarrow G}(n,t)$ and $\lambda^1_{D \rightarrow G}(n,t) \geq 0$, and that the same properties hold for $\Lambda$. Also we assume that they are bounded and continuous as function of time.\\
The reference player would like to choose its transition rate $\lambda_{D \rightarrow G}$, possibly different from $\Lambda$, in order to minimize
\begin{equation*}
\begin{split}
    v_n^D(t,\Lambda,\lambda) &= \mathbb{E}_{A_t(D,n)}^{\Lambda, \lambda}\left[\int_{t}^{T}c\left(X_s^{1}, \frac{N_s}{N}, \lambda(s)\right)\,ds\right] \\
                             &= \mathbb{E}_{A_t(D,n)}^{\Lambda, \lambda}\left[\int_{t}^{T} \left(\sigma^2 \lambda_{D \rightarrow G}^2(n,s) + \frac{N_G(X_s^N)}{N}\right)1_{\{X_s^{1}=D\}}\,ds\right]
\end{split}
\end{equation*}
where the subscript $A_t(D,n)$ means we are considering the expectation conditioned on $X_t=D$ and $N_t = n$. That is, the reference player looks for the control $\lambda$ which is a solution to the following minimization problem
\begin{equation*}
    v_n^D(t;\Lambda) = \inf_{\lambda} v_n^{D}(t,\Lambda,\lambda),
\end{equation*}
where the minimization is performed over the set of all admissible controls $\lambda$. In particular, the function $v_n^D(t;\Lambda)$ is the value function for the reference player associated to the strategy $\Lambda$ of the remaining $N$ players. The control that attains the minimum above can be called the best response (for the reference player) to a control $\Lambda$.\\
\indent We conclude this section with the following important remarks:
\begin{itemize}
\item[(i)] In the Markov model (without control) in Section \ref{sec:Markov}, with Green-transition rate%
\[
\lambda_{N}\left(x_N,t\right)  =a\left(  t\right)  \frac{N_{G}\left(x_N\right)  }{N}%
\]
subjects react time-by-time, based on \textit{observed environment }%
$\frac{N_{G}\left(x_N\right)  }{N}$.
\item[(ii)] In the game model with payoff, agents are asked to make a decision at
the \textit{same initial time} depending on an \textit{unknown
future environment }$\frac{N_{G}(X_s^N)}{N}$.
\item[(iii)] In the first case, the consequence is a \textit{slow (but exponential)
adaptation} to the environment.
\item[(iv)] Opposite, in the second case, the consequence is a \textit{fast
reaction} to\ a potential environment.
\end{itemize}}

\subsection{Simulations}

Simulating the \textcolor{black}{mean-field Nash equilibrium in Theorem \ref{th:nash} is less
straightforward than a usual initial value problem since from the ODE point of view this problem is somewhat non-standard as variable has initial conditions whereas the other variable have prescribed terminal data}. We have used a shooting
method, namely looking for the initial condition of the variable
$\widetilde{v}_{D}\left(  t\right)  $ which, along with the initial condition
$p_{0}$, solving forward the system, we get $\widetilde{v}_{D}\left(
T\right)  =0$.

\begin{figure}
\centerline{\includegraphics[width=3.8in]{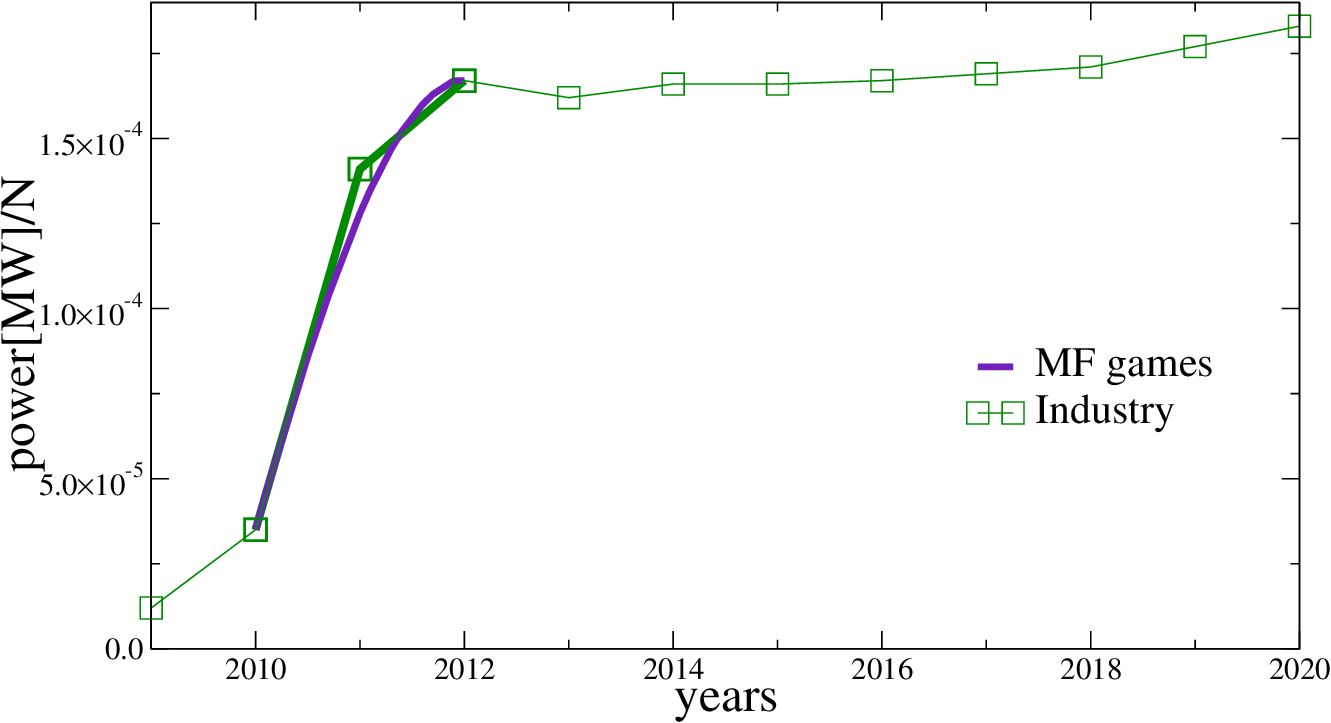}}
\vspace*{8pt}
\caption{Sudden increase in 2011 of the Industrial compartment.}\label{fig:games}
\end{figure}

The result, illustrated in Figure \ref{fig:games} is that the curve $\widetilde{p}%
_{G}\left(  t\right)  $ is concave. Namely, the result is completely different
from the exponential growth of the Markov model. Tuning the parameters, it
may fit the data, as shown in the figure. %

Let us stress the intuitive reason for the concavity, opposite to exponential
growth: it stands in the intimate structure of a game, opposite to a Markov
mechanism. In a game, agents have a tendency to act in advance, to anticipate
the move of the other players. The reason is the limited amount of subsidies,
which does not allow to get them if they are exhausted by other players
before. This is the opposite of procrastination, in a sense.

\section{Conclusions\label{sect Conclusions}}
\textcolor{black}{The conclusion drawn from this analysis, considering the effectiveness of the 2011 governmental action, is that a similar structure should be applied to the domestic sector in the future to stimulate a stronger increase in domestic photovoltaic usage, which is currently stagnating. To motivate homeowners to take action, it is essential to enhance their planning abilities and prepare for a defined starting point, ensuring that individuals are aware of this moment. This can potentially be achieved through targeted advertisements and guidance for homeowners.
Additionally, a more comprehensive approach could involve enhancing consumers' ability to make informed decisions, or ``deliberate" effectively. Educational initiatives aimed at this objective could be offered by public authorities in various formats, such as free cultural and educational programs for adults or engaging activities for school-aged children.
However, it is important to be cautious against simplistic recommendations. A comparison of the time series data from Italy and Germany (see Figure 1 in Section \ref{sect time-series}) reveals that both countries experienced two periods of exponential growth at different rates, with Italy experiencing a significant surge around 2011. Ultimately, while both the rate of exponential increase and the imitation coefficient \( a(t) \) are crucial, it remains a question for further research whether it is more effective to focus on increasing the coefficient or to trigger a dramatic event similar to the one in 2011.}

\newpage
\appendix
\section{A brief overview on the development of the photovoltaic systems
in Italy\label{sect development Italy}}
In this subsection, \textcolor{black}{for the sake of completeness}, we briefly review the initiatives implemented by the
Italian government encouraged the diffusion of solar photovoltaic systems (PVs, henceforth) from 2005 until today. These initiatives are called  ``Conto Energia'' (CE); each CE guarantees contracts with fixed conditions for 20 years for grid-connected PVs with at least 1kW of peak power. Local electricity providers are required by law to buy the electricity that is generated by PVs. The first CE started in 2005, and it was a net metering plan (``scambio sul posto'') designed for small PVs. The plan was meant to favor the direct use of self-produced electricity. Besides payment for each produced kWh of electricity, the consumer received additional rewards for directly consuming the self-generated energy. The CE2 was available to all PVs, but it was designed for larger plants with no or limited direct electricity self-consumption. The electricity produced was sold to the local energy supplier, for which the CE guarantees an additional FiT. It is essential to mention that in
each new version of the CE, the FiT was decreased (from 0.36 \euro/kWh in
2006 to 0.20 \euro/kWh in 2012). With the introduction of CE4 (2011) direct consumption was rewarded financially. The CE5, unlike the CE4, provided incentives based on the energy fed into the grid and a premium rate for self-consumed energy.

After the end of the fifth CE program, FiT and premium schemes were
dropped, and a tax credit program was implemented in 2013. After six years, in 2019, a new incentive decree for photovoltaic systems (RES1) was reintroduced, reserved for systems with a more than 20 kW capacity but not more than 1 MW. Subsidies are paid based on net electricity produced and fed into the grid. The unit incentive varies according to the size of the plant. An incentive is provided for plants that replace asbestos or eternity roofing, and a bonus on self-consumption of energy (provided it is more significant than 40\% and the building is on a roof) is issued. For residential customers, a subsidized tax deduction is set at 50\% instead of 36\%.

However, in May 2020, the Italian government issued the ``Revival Decree'' (Decree Law 34/2020), introducing a further
increase to 110\%. Depending on whether the installation is
connected to energy-saving measures or not, the 110\% tax deduction can be
applied to the entire investment (max 2400 \euro/kW) or only to a part of it (max 1600 \euro/kW). In addition, the energy not consumed directly is transferred free of charge to the grid. Moreover, the Revival Decree provides for the subsidized tax at 110\% and also for the implementation of battery energy storage systems up to an amount of 1000 \euro/kWh. In addition, the Relaunch Decree provides for the subsidized tax at 110\% and the implementation of storage systems up to 1000 \euro/kWh.

Finally, an additional policy measure was introduced by the Ministerial
Decree of September 16, 2020, which provides incentives for the configuration of collective self-consumption and renewable energy communities equal to 100 \euro/MWh and 110 \euro/MWh, respectively. The incentive lasts for 20 years, does not apply to plants exceeding a power of 200 kW, and has a duration of 60 days from the entry into force of the decree.

\textcolor{black}{\section{Dynkin's formula and Aldous criterium\label{dynkin and Aldous}}
This section reviews some aspects of non-homogeneous continuous-time Markov processes.  We start this presentation by considering a non-homogeneous Markov process in continuous-time with finite state-space $\mathcal{S}$. Then, we will consider a sequence of Markov processes $(X_t^N)_{t \geq 0}$ on state-spaces $\mathcal{S}_N$: this is the case analyzed in the present article.\\ \\ \indent Let $(X_t)_{t \geq 0}$ be a non-homogeneous continuous-time Markov process with finite state-space $\mathcal{S}$. It is described by the time-dependent transition rates $\lambda(x, y, t)$ from a state $x$ to a state $y$ in $\mathcal{S}$ and the associated infinitesimal generator
\begin{equation*}
    (\mathcal{L}_t F)(x) = \sum_{\substack{y \neq x \\ x, y \in \mathcal{S}}} \lambda(x,y,t)(F(y)-F(x)),
\end{equation*}
with $F$ a test function.  Now,    denote by $(X_t^{x})_{t \geq 0}$ the Markov process indexed by the initial condition $x$ at time zero.  The content of Dynkin's formulas in this particular framework -- the result holds in higher generality, which, however, is not needed here -- is given by the fact that for every test function $F:\mathcal{S}\rightarrow\mathbb{R}$ the following two processes $(M_t^{F})$ and $(N_t^{F})$ are martingales:
\begin{equation}\label{eq::martingales}
    \begin{split}
        M_t^F &:= F(X_t^{x})-F(X_t^{0})-\int_{0}^{t}(\mathcal{L}_s F)(X_s^{x})\,ds,\\
        N_t^F &:= (M_t^F)^2 - \int_{0}^{t}(\mathcal{L}_s F^2 - 2 F \mathcal{L}_s F)(X_s^{x})\,ds;
    \end{split}
\end{equation}
see \cite{KL}, Appendix 1, Section 5\footnote{Notice that the theory presented in \cite{KL} is for homogeneous continuous-time Markov processes. However, starting from it, one can derive analogous results for the non-homogeneous case.}. The quantities in Equation \eqref{eq::martingales} are defined as
\begin{equation*}
    \begin{split}
        &(\mathcal{L}_s F^2 - 2 F \mathcal{L}_s F)(x) = \sum_{\substack{y \neq x \\ x, y \in \mathcal{S}}} \lambda(x,y,s)(F(y)-F(x))^2\\
        \text{where\,\,\,}&(\mathcal{L}_s F^2)(x) = \sum_{\substack{y \neq x \\ x, y \in \mathcal{S}}} \lambda(x,y,s)(F^2(y)-F^2(x)),\\
                          &2 F(x) (\mathcal{L}_s F)(x) = \sum_{\substack{y \neq x \\ x, y \in \mathcal{S}}}  \lambda(x,y,s) (2 F(x) F(y) - 2 F^2(x)).
    \end{split}
\end{equation*}
In particular, the difference between the last two equations is indeed given by $\sum_{\substack{y \neq x \\ x, y \in \mathcal{S}}}\lambda(x,y,s)(F(y)-F(x))^2$ and, being the average of $M_t^F$ and $N_t^F$ equal to zero, Equation \eqref{eq::martingales} implies:
\begin{equation}\label{eq::firstandsecond}
    \begin{split}
        \mathbb{E}[F(X_t^{x})] &= \mathbb{E}[F(X_0^{x})] + \mathbb{E}\left[\int_{0}^{t}(\mathcal{L}_s F)(X_s^{x})\,ds,\right]\\
        \mathbb{E}[(M_t^F)^2] &= \mathbb{E}\left[\int_{0}^{t}(\mathcal{L}_s F^2 - 2 F \mathcal{L}_s F)(X_s^{x})\,ds\right].
    \end{split}
\end{equation}
\noindent   The first equation is the so-called first Dynkin's formula, whereas the second one is the so-called second Dynkin's formula.\\ 
\indent Instead, when we have a sequence of Markov processes $(X_t^N)_{t \geq 0}$ on state spaces $\mathcal{S}_N$ with generators $\mathcal{L}_{t,N}$ and we are interested in the convergence of a real-valued summary $F_N(X_t^N)$ with $(F_N)_{N \in \mathbb{N}}$ a sequence of functions, $F_N:\mathcal{S}_N\rightarrow\mathbb{R}$, Aldous criteria is more practical; see \cite{KL}, Chapter 4, Proposition 1.6.   In what follows, we give for granted the definition of Skorohod space $\mathcal{D}(0,T)$ (see, e.g., \cite{KL}). Aldous criteria -- particularized to our case -- states that the laws $P_N$ of $F_N(X^{N}_{\cdot})$ on Borel sets of $\mathcal{D}(0,T)$ are tight if the following condition holds: for every $\epsilon>0$ we have that
\begin{equation}\label{eq::aldous}
    \lim_{\gamma \rightarrow 0}\limsup_{N \rightarrow \infty} \sup_{\substack{\tau \in \Psi \\ \theta \in [0,\gamma]}}\mathbb{P}(|F_N(X_{\tau+\theta}^N)-F_N(X_{\tau}^N)|>\epsilon)=0,
\end{equation}
where $\Psi$ is the family of stopping times less or equal than $T$, and where, in order not to burden the notation, we omitted the dependence on the initial condition. By the first Dynkin's formula we have that
\begin{equation*}
    F_N(X^N_{\tau+\theta})-F_N(X^N_{\tau}) = \int_{\tau}^{\tau+\theta}(\mathcal{L}_{s,N}F_N)(X_s^{N})\,ds + M_{\tau+\theta}^{F_N}-M_{\tau}^{F_N},
\end{equation*}
where $M_{\tau}^{F_N}$ is a martingale. Since
\begin{equation*}
    \begin{split}
        \mathbb{P}(|F_N(X_{\tau+\theta}^N)-F_N(X_{\tau}^N)|>\epsilon) &\leq \mathbb{P}\left(\left|\int_{\tau}^{\tau+\theta}(\mathcal{L}_{s,N}F_N)(X_s^{N})\,ds\right|>\frac{\epsilon}{2}\right)\\
        &+\mathbb{P}\left(\left|M_{\tau+\theta}^{F_N}-M_{\tau}^{F_N}\right|>\frac{\epsilon}{2}\right),
    \end{split}
\end{equation*}
it is sufficient to prove the analog of condition \eqref{eq::aldous} for each one of the two terms on the right-hand side of the previous inequality.   In many cases it happens that there exists a constant $C_1>0$ such that
\begin{equation}\label{eq::cond1}
    \sup_{\substack{s \in [0,T] \\ x_N \in \mathcal{S}_N}}\left|(\mathcal{L}_{s,N}F_N)(x_N)\right| \leq C_1,
\end{equation}
so that $\left|\int_{\tau}^{\tau+\theta}(\mathcal{L}_{s,N}F_N)(X_s^{N})\,ds\right|\leq C\theta$. Whence, for $\gamma$ small enough it happens that $\theta \in [0,\gamma]$ satisfies $C\theta \leq \frac{\epsilon}{2}$; therefore $\mathbb{P}\left(\left|\int_{\tau}^{\tau+\theta}(\mathcal{L}_{s,N}F_N)(X_s^{N})\,ds\right|>\frac{\epsilon}{2}\right)=0$.  It remains to bound the second term. We have:
\begin{equation*}
    \begin{split}
        &\mathbb{P}\left(\left|M_{\tau+\theta}^{F_N}-M_{\tau}^{F_N}\right|>\frac{\epsilon}{2}\right) \leq \frac{4}{\epsilon^2}\mathbb{E}\left[\left|M_{\tau+\theta}^{F_N}-M_{\tau}^{F_N}\right|^2\right]\\
        &= \frac{4}{\epsilon^2}\left(\mathbb{E}\left[\left|M_{\tau+\theta}^{F_N}\right|^2-\left|M_{\tau}^{F_N}\right|^2\right]\right)\\
        &=\frac{4}{\epsilon^2}\mathbb{E}\left[\int_{\tau}^{\tau+\theta}(\mathcal{L}_{s,N}F_N^2-2 F_N \mathcal{L}_{s,N}F_N)(X_s^{N})\,ds\right],
    \end{split}
\end{equation*}
where in the first equality we have used the martingale properties with respect to the $\sigma$-algebra associated to $\tau$, whereas in the second one the second Dyinkin's formula.   Again, in many cases it happens that there exists a constant $C_2>0$ such that
\begin{equation}\label{eq::cond2}
    \sup_{\substack{s \in [0,T] \\ x_N \in \mathcal{S}_N}}\left|(\mathcal{L}_{s,N}F_N^2-2 F_N \mathcal{L}_{s,N}F_N)(x_N)\right|\leq C_2.
\end{equation}
In this case we have that 
\begin{equation*}
    \mathbb{P}\left(\left|M_{\tau+\theta}^{F_N}-M_{\tau}^{F_N}\right|>\frac{\epsilon}{2}\right) \leq \frac{4 C_2}{\epsilon^2}\theta,
\end{equation*}
and therefore the limit in Equation \eqref{eq::aldous} is equal to zero. Summarizing, as a consequence of Aldous condition, we have the following
\begin{proposition}\label{prop::tightness}
    Suppose that conditions \eqref{eq::cond1} and \eqref{eq::cond1} hold.   Then the laws $P_N$ of $F_N(X^N_{\cdot})$ on Borel sets of $\mathcal{D}(0,T)$ are tight. 
\end{proposition}}

\section*{Declarations}
\noindent {\bf Conflict of interest.} The authors declare that they have no conflict of
interest.

\section*{Acknowledgment}
The paper constitutes the first result of the interdisciplinary and inter-institutional research group working within the research project funded by the Italian Ministry of University and Research under the title: ``Teorie e strumenti per la transizione ecologica: profili filosofici, matematici, etici e giuridici relativi alla sfida di sostenibilità del carbon budget / Theories and tools for the ecological transition: philosophical, mathematical, ethical and juridical profiles related to the sustainability challenge of the carbon budget'' (Prog. MUR-PRO3 2022-2024).
The project gathers three research units, respectively based at the Scuola Normale Superiore, the Scuola Superiore Sant'Anna (coord.) and the University School for Advanced Studies - IUSS Pavia. This paper is the result of the shared work by Scuola Normale Superiore and Scuola Superiore Sant'Anna research units.
S.M. acknowledges partial support by Italian Ministry of University and Research under the PRIN project Convergence and Stability of Reaction and Interaction Network Dynamics (ConStRAINeD) - 2022XRWY7W. 

It is a pleasure to thank GSE for providing us the data used in this work and, in particular, we thank Dr. Paolo Liberatore for his helpful clarifications.

\section*{Author Contributions}
Conceptualization (all authors), Methodology (ML, GL, SM, FF), Writing (all authors), Mathematical investigation (ML, GL, SM, FF), Data analysis (SM, FF).

\bibliographystyle{alpha}

\newcommand{\etalchar}[1]{$^{#1}$}

\end{document}